\documentclass{svproc}
\usepackage{url}
\usepackage{amsmath}
\usepackage{amsfonts}
\usepackage{bm}

\newcommand{\bra}[1]{\langle \, #1 \, |}

\newcommand{\kket}[1]{\, #1 \, \rangle}
\newcommand{\eq}{\text{equiv}}

\begin{document}
\mainmatter              % start of a contribution
\title{Local meson-baryon coupled-channels potential for the $\Lambda(1405)$}
\titlerunning{Local meson-baryon coupled-channels potential for the $\Lambda(1405)$}  % abbreviated title (for running head)
\author{Kenta Miyahara\inst{1} \and Tetsuo Hyodo\inst{2}
 \and Wolfram~Weise\inst{3}}
\authorrunning{Kenta Miyahara et al.} % abbreviated author list (for running head)
\tocauthor{Kenta Miyahara, Tetsuo Hyodo, and Wolfram Weise}
\institute{Department of Physics, Graduate School of Science, Kyoto University, Kyoto 606-8502, Japan,
\and
Yukawa Institute for Theoretical Physics, Kyoto University, Kyoto 606-8502, Japan\\
\email{hyodo@yukawa.kyoto-u.ac.jp}
\and
Physik-Department, Technische Universit\"at M\"unchen, 85748 Garching, Germany
}

\maketitle              % typeset the title of the contribution

\begin{abstract}
A local coupled-channels $\bar{K}N$-$\pi\Sigma$-$\pi\Lambda$ potential is constructed, equivalently reproducing the scattering amplitude in chiral SU(3) dynamics. By analyzing the wave function of the $\Lambda(1405)$, we show the $\bar{K}N$ dominance of the upper pole within the two-pole structure of the $\Lambda(1405)$. The resulting potential will be useful for investigating $\bar{K}$ few-nucleon systems including their coupled channels.
%The abstract should summarize the contents of the paper
%using at least 70 and at most 150 words. It will be set in 9-point
%font size and be inset 1.0 cm from the right and left margins.
%There will be two blank lines before and after the Abstract. \dots
% We would like to encourage you to list your keywords within
% the abstract section using the \keywords{...} command.
\end{abstract}
\section{Introduction}
Systems of an antikaon and few nucleons are of considerable interest in few-body physics because the low-energy $\bar{K}N$ interaction is considered to be sufficiently attractive to generate a quasibound state, the $\Lambda(1405)$, below $\bar{K}N$ threshold~\cite{Dalitz:1959dn,Dalitz:1960du,Dalitz:1967fp}. One of the recent theoretical achievements is the determination of the precise meson-baryon scattering amplitude, including the pole position of the $\Lambda(1405)$~\cite{Tanabashi:2018oca}. Currently, a whole set of experimental data near the $\bar{K}N$ threshold is successfully described in the framework of chiral SU(3) coupled-channels dynamics~\cite{Ikeda:2011pi,Ikeda:2012au}.

In order to study the $\bar{K}$ few-nucleon systems with established techniques for rigorous few-body calculations, it is desirable to construct a local meson-baryon potential. A detailed strategy for constructing such a local potential equivalent to chiral SU(3) dynamics has been developed in Ref.~\cite{Hyodo:2007jq}. It is also shown that one can renormalize the effects of $\pi Y$ ($\equiv\pi\Sigma, \pi\Lambda$) channels to obtain a single-channel $\bar{K}N$ potential. By applying this strategy to the scattering amplitude in Refs.~\cite{Ikeda:2011pi,Ikeda:2012au}, an elaborate single-channel $\bar{K}N$ potential has been derived in Ref.~\cite{Miyahara:2015bya}. This potential is applied to the investigations of few-body kaonic nuclei~\cite{Ohnishi:2017uni} and kaonic deuterium~\cite{Hoshino:2017mty}.

While the single-channel $\bar{K}N$ potential exactly reproduces the two-body scattering amplitude, in the application to the few-body systems, dynamical effects of the coupled-channels (such as $\pi Y$ plus nucleons) are not properly included. For states near the $\bar{K}N$ threshold, the neglect of such components does not affect observables very much. However, the results in Ref.~\cite{Ohnishi:2017uni} suggest that the quasibound state of $\bar{K}$ and six nucleons appears near the threshold of $\pi \Sigma$ plus five nucleons. In such cases, the explicit treatment of coupled-channels dynamics is expected to give sizable contributions.

Here we construct the coupled-channels $\bar{K}N$-$\pi\Sigma$-$\pi\Lambda$ potential~\cite{Miyahara:2018onh} equivalent to the scattering amplitude in Refs.~\cite{Ikeda:2011pi,Ikeda:2012au}. As an application within the two-body sector, we study the meson-baryon fractions in the $\Lambda(1405)$ by analyzing the wave function of the quasibound state.

\section{Coupled-channels meson-baryon potential}
We start from the coupled-channels Schr\"odinger equation for a nonrelativistic energy $E$,
\begin{align}
\left[ -\frac{\nabla^2}{2\mu_i}\delta_{ij} + \Delta M_i\, \delta_{ij} + V^\eq_{ij}(\bm{r},E) \right]\psi_{j}(\bm{r}) = E \psi_{i}(\bm{r})~,
\label{eq:Sch_psi}
%\eqref{eq:Sch_psi}
\end{align}
where $\mu_{i}$, $\Delta M_{i}$, and $\psi_{i}$ are the reduced mass, the energy difference from the reference threshold, and the wave function in channel $i$, respectively. The coupled-channels potential $V_{ij}^{\eq}$ has a matrix form in channel basis with $i=\bar{K}N$, $\pi\Sigma$, and $\pi\Lambda$. In contrast to the single-channel $\bar{K}N$ potential which has a complex strength
%where the potential strength is complex 
due to the absorption into the $\pi Y$ channels, the strengths of $V_{ij}^{\eq}$ are real valued. In addition, $V_{ij}^{\eq}$ has an energy dependence, which originates from the energy dependence of the chiral interaction and from the renormalization of higher energy channels such as $\eta\Lambda$ and $K\Xi$.

Our task is to determine $V^\eq_{ij}$ such that the scattering amplitude from Eq.~\eqref{eq:Sch_psi} reproduces the one in Refs.~\cite{Ikeda:2011pi,Ikeda:2012au}. The potential $V^\eq_{ij}$ should be related to the interaction kernel in chiral SU(3) dynamics, but there is no direct way of converting the interaction kernel, given that the framework is different. We therefore introduce several matching conditions of $V^\eq_{ij}$ and the interaction kernel, not only on the real energy axis but also in the complex energy plane, in order to systematically determine $V^\eq_{ij}$. For the details of the potential construction procedure, see Ref.~\cite{Miyahara:2018onh}. 

For practical applications, it is useful to represent $V_{ij}^{\eq}$ in a conveniently parametrized form. We use the following parametrization:
\begin{align}
V_{ij}^\eq(\bm{r},E) &= e^{-r^2(1/2b_{i}^2+1/2b_{j}^2)}
\sum_{\alpha=0}^{\alpha_{\rm max}} K_{\alpha,{ij}}\,
\left(\frac{E}{100 \text{ MeV}}\right)^\alpha\, . 
\label{eq:Vfit} 
%\eqref{eq:Vfit}
\end{align}
The spatial distribution is assumed to be a Gaussian form with the range parameter $b_{i}$. The energy dependence is parametrized by a polynomial of order $\alpha_{\rm max}$ with real coefficients $K_{\alpha,{ij}}$. These parameters are determined by the matching conditions mentioned above, and the results are tabulated in Ref.~\cite{Miyahara:2018onh}. One remarkable fact is that the energy dependence of $V_{ij}^\eq$ can be accurately parametrized by a second-order polynomial ($\alpha_{\rm max}=2$). This is in sharp contrast to the single-channel version in Ref.~\cite{Miyahara:2018onh} where one needs $\alpha_{\rm max}$ as large as 10 for sufficient accuracy. In other words, the explicit inclusion of the $\pi Y$ channels permits a parametrization of the potential strength with a more natural, much weaker energy dependence. 

\section{Compositeness of the $\Lambda(1405)$}
In general, a resonance eigenstate is expressed by a pole of the scattering amplitude in the complex energy plane. In the case of the $\Lambda(1405)$, the resonance is expressed by two poles~\cite{Jido:2003cb,Tanabashi:2018oca}. From the coupled-channels potential, we find poles at
%\begin{align}
%\sqrt{s} &= 1424-27i \text{ MeV} ,\quad
%\sqrt{s} = 1380-81i \text{ MeV} ,
%\end{align}
$\sqrt{s} = 1424-27i$ MeV and $\sqrt{s} = 1380-81i$
with $\sqrt{s}=E+M_{N}+m_{K}$, consistently with the original scattering amplitude~\cite{Ikeda:2011pi,Ikeda:2012au}. The wave function of the $\Lambda(1405)$ can be obtained by evaluating Eq.\,\eqref{eq:Sch_psi} at these pole energies. Starting from the wave function it is possible to extract the properties of the resonance. It should be kept in mind that due to the unstable nature of resonances, the expectation value of operators (such as norm of the state) becomes complex.

Here we discuss the compositeness $X_{i}$ and the ``elementarity'' $Z$~\cite{Hyodo:2013nka} which are related to the norm of the wave function in channel $i$:
\begin{align}
  X_{i}&=\bra{\psi^{\dag}_{i}}\kket{\psi_{i}} , \quad 
  Z=1-\sum_{i}X_{i},
\end{align}
where $\bra{\psi^{\dag}_{i}}$ is a left eigenstate $\bra{\psi^{\dag}_{i}}H=\bra{\psi^{\dag}_{i}}E$. When we deal with resonances, $\bra{\psi_{i}^{\dag}}$ is different from $\bra{\psi_{i}}$. The compositeness $X_{i}$ represents the fraction of the channel $i$ component in the total wave function. When the potential is energy dependent, the  sum of $X_{i}$s does not become unity, and the elementarity $Z$ is introduced. Using the Feshbach projection method, $Z$ can be expressed by the expectation value of the energy derivative of the potential~\cite{Miyahara:2018onh}. Originally, $Z$ has been introduced as a field renormalization constant of the bare state representing the elementary component of the composite system~\cite{Weinberg:1965zz}. In the present context, it is understood as the contributions not explicitly included in the model space.

\begin{table}[b]
\caption{Compositeness $X_{i}$ and elementarity $Z$ for high-mass pole of the $\Lambda(1405)$ coupled-channels system.}
\label{tbl:compositeness}
\begin{center}
\begin{tabular}{lccc}
\hline
Method & $X_{\pi\Sigma}$ & $X_{\bar{K} N}$ & $Z$\\[2pt]
\hline
Coupled-channels potential & $-0.02-0.25i$ & \ $1.01-0.13i$ \ & $0.01+0.37i$ \\
Single-channels potential~\cite{Miyahara:2015bya}  &               & \ $1.01-0.07i$ \ &  \\
Residue of the pole~\cite{Sekihara:2014kya} & $-0.19-0.22i$ & \ $1.14+0.01i$ \ & $0.05+0.21i$ \\[2pt]
\hline
\end{tabular}
\end{center}
\end{table}

We show the results of $X_{\pi\Sigma}$, $X_{\bar{K}N}$, and $Z$ for high-mass pole at $1424-27i$ MeV in Table~\ref{tbl:compositeness}. The results are consistent with those obtained by the single-channel potential\,\cite{Miyahara:2015bya}, as well as with the evaluation by the residue of the pole~\cite{Sekihara:2014kya}. In all cases, $X_{\bar{K}N}$ is close to unity and the others are almost zero. Hence, we conclude the $\bar{K}N$ dominance of the high-mass pole of the $\Lambda(1405)$.

\section{Summary}
We have constructed a coupled-channels meson-baryon ($\bar{K}N\leftrightarrow \pi Y$) potential based on chiral SU(3) dynamics. The evaluation of the compositeness tells us that the high-mass pole of the $\Lambda(1405)$ is dominated by the $\bar{K}N$ component. This potential will be useful in order to shed new light on meson-baryon few-body systems with strangeness.


\end{document}